\documentclass[twoside]{article}

\usepackage{lipsum} 
\usepackage{aas_macros}

\usepackage[sc]{mathpazo} 
\usepackage[T1]{fontenc} 
\usepackage{microtype} 

\usepackage[hmarginratio=1:1,top=32mm,columnsep=20pt]{geometry} 
\usepackage{multicol} 
\usepackage[hang, small,labelfont=bf,up,textfont=it,up]{caption} 
\usepackage{booktabs} 
\usepackage{float} 
\usepackage{hyperref} 
\usepackage{natbib} 
\usepackage{graphicx}
\usepackage{caption} 
\usepackage{subcaption} 
\usepackage{lettrine} 
\usepackage{paralist} 

\usepackage{abstract} 

\usepackage{titlesec} 
\renewcommand\thesection{\Roman{section}} 
\renewcommand\thesubsection{\Roman{subsection}} 
\titleformat{\section}[block]{\large\scshape\centering}{\thesection.}{1em}{} 
\titleformat{\subsection}[block]{\large}{\thesubsection.}{1em}{} 

\usepackage{fancyhdr} 
\title{\vspace{-15mm}\fontsize{24pt}{10pt}\selectfont\textbf{Independent Discovery of a Sub-Earth in the Habitable Zone Around a Very Close Solar-Mass Star}} 

\author{
\large
\textsc{Michael B. Lund$^1$, Robert J. Siverd$^2$, and Ponder Stibbons$^3$}\\
\normalsize $^1$Vanderbilt University\\
\normalsize $^2$Las Cumbres Observatory \\
\normalsize $^3$Unseen University, Ankh-Morpok \\ 
\normalsize \href{mailto:michael.b.lund@vanderbilt.edu}{michael.b.lund@vanderbilt.edu} 
\vspace{-5mm}
}
\date{}

\begin{document}

\maketitle 

\thispagestyle{fancy} 


\begin{abstract}

\noindent With the wealth of planets that have been discovered over the past $\sim$ 20 years, the field can broadly be divided into two regimes. For understanding broad occurrence and formation rates, large numbers of planets allow for population statistics to be calculated, and this work preferentially tends towards fainter planets (and fainter host stars) to allow for a large number of detections. The second regime is the detailed understanding of a single planet, with particular consideration to planetary structure and atmosphere, and in this case benefits from finding individual planets (and host stars) that are very close, and subsequently, very bright. The closest of these also provide very novel possibilities for exploration if they are close enough that travel time to them is relatively low, something that would be extremely unlikely for more distant planets. Here, we announce the independent discovery of a sub-earth planet orbiting in the habitable zone of a very close solar-mass star using a novel processing technique and observations from the Kilodegree Extremely Little Telescope (KELT).

\end{abstract}


\begin{multicols}{2} 

\section{Introduction}
\lettrine[nindent=0em,lines=3]{T} he search for planets can be divided into the attempt to find individual planets that provide unique insight into planets (either by being favorable to in-depth characterization or being intrinsically unusual planets) and the attempt to identify large numbers of planets that can be used for population statistics. While the latter has made significant progress with the large number of planets provided by the \emph{Kepler} mission, individual planets well-suited to characterization have been more difficult to find due to the need for these planets to be both nearby and optimial for observation. For example, \citet{Dressing2015} noted that the nearest habitable zone planets around red dwarfs would be 2.6 or 10.6 parsecs away, with the latter including an additional constraint that the planet be transiting its host star. The inherent rarity of nearby habitable planets has meant that any nearby planets have been significant discoveries, such as the Gliese 876 system \citep{Marcy1998, Delfosse1998}, Formalhaut b \citep{Kalas2008}, and Proxima Centauri b \citep{2016Natur.536..437A}. Consequentially, the discovery of any planet that is closer than these and thereby well-suited to additional observations would be an extremely valuable contribution to the understanding of planets, and the discovery of a planet in the habitable zone would be particularly useful.

In this paper we present a planet with a period of $\sim$700 days around an extremely close G2V host star, which also represents and independent discovery of the planet that was recently announced in announcements by The Astronomer's Telegram \citep{2018ATel11448....1D, 2018ATel11449....1D}.

\section{Detection}
\subsection{Kilodegree Extremely Little Telescope (KELT)}
The Kilodegree Extremely Little Telescope (KELT)-North survey has been continuously observing areas in the northern hemisphere since 2006 for transiting Hot Jupiters around bright stars (8<V<10) from its location at Winer Observatory in Arizona. A single KELT-North field is 26deg by 26 deg with a 4k x 4k CCD yielding a 23'' pixel scale. Observations cover the entire visible sky in a broad R-band filter at a cadence of 15-30 minutes \citep{Pepper2007, Pepper2012}. The standard data reduction steps taken for KELT images is further discussed in \citet{Siverd2012}. In this paper we use observations taken from  KELT-North field 05, which is centered on $\alpha$ =  07hr 50m 24.0s, $\delta$ = $+31^{\circ}$ 39$\prime$  56$\prime\prime$ J2000.

\subsection{Photometry}
\begin{figure*}[!htb]
  \begin{center}
   \includegraphics[width=\textwidth]{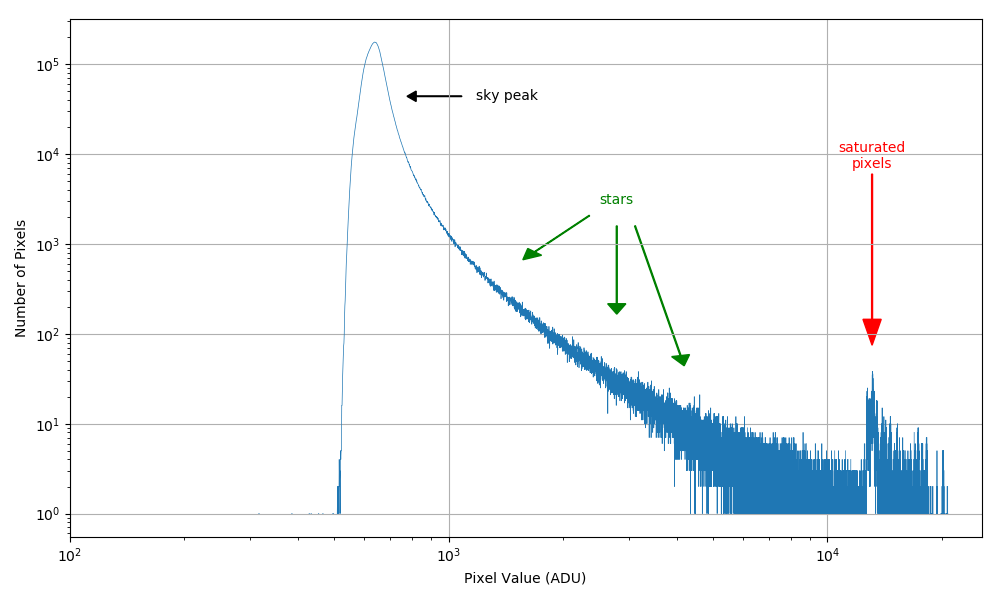}
  \end{center}
  \caption{A histogram of a typical KELT image after dark subtraction and flat-fielding. There is a single broad and high peak due to the sky and a long, shallow tail caused by stars. There is also a small peak near 16000 ADU (red arrow) caused by a pile-up at the saturation value. Our detection statistic reports the fraction of pixels with a value above 16000 ADU, i.e., what fraction of pixels live in this "bump" marked by the red arrow. }
  \label{fig:hist}
\end{figure*}

\begin{figure*}[!htb]
  \begin{center}
   \includegraphics[width=\textwidth]{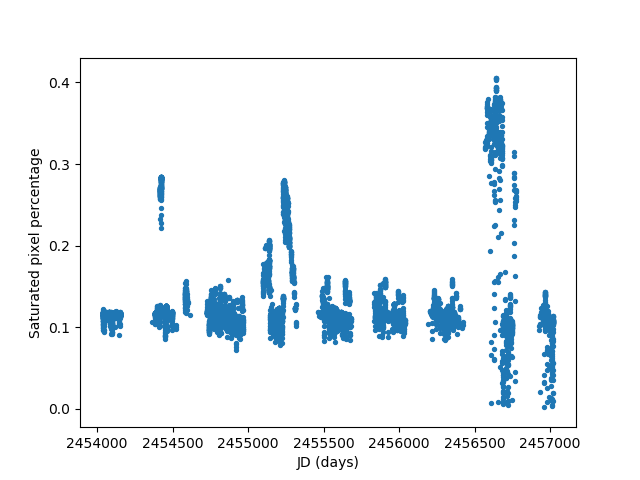}
  \end{center}
  \caption{The percentage of pixels that are saturated as a function of time, serving as a crude light curve. The time period examined is about 8 years in length. In typical images, roughly 0.1\% of pixels are saturated. Periods where the number increases significantly indicate the presence of one or more extremely bright transients in the field. Events represent times where significantly brighter objects were located in the field.}
  \label{fig:lc}
\end{figure*}

To search for a particularly bright object with unknown position, we adopt a novel technique by analyzing the number of saturated pixels in each image. Our test statistic is simply the percentage of pixels in an uncalibrated image with value > 16000 ADU\footnote{The KELT-North CCD operates in 14-bit readout mode. This causes ADC saturation to occur at 16383 ADU rather than the usual 65535 expected for a 16-bit detector}. This statistic offers great sensitivity to extremely bright transients. This enhanced sensitivity is caused by detector blooming: although many saturated pixels are present in typical images, a single extremely bright source may generate up to a few x $10^4$ saturated pixels ($\sim10$ columns $\times$ 4096 pixels per column), producing a significant change in the total number of saturated pixels. Since the number of saturated pixels tends to be small (roughly 0.1\% for KN05) and a typical image histogram has a shallow slope (see Fig~\ref{fig:hist}), the saturated pixel fraction is much more sensitive to the presence of an extremely bright transient than it is to changes in mean image level. To perform our analysis, we produced a "light curve" of the KN05 field consisting of mid-exposure JD (UTC) and the value of this statistic. This time series is illustrated in Fig~\ref{fig:lc}.

\begin{figure*}[!htb]
  \begin{center}
   \includegraphics[width=\textwidth]{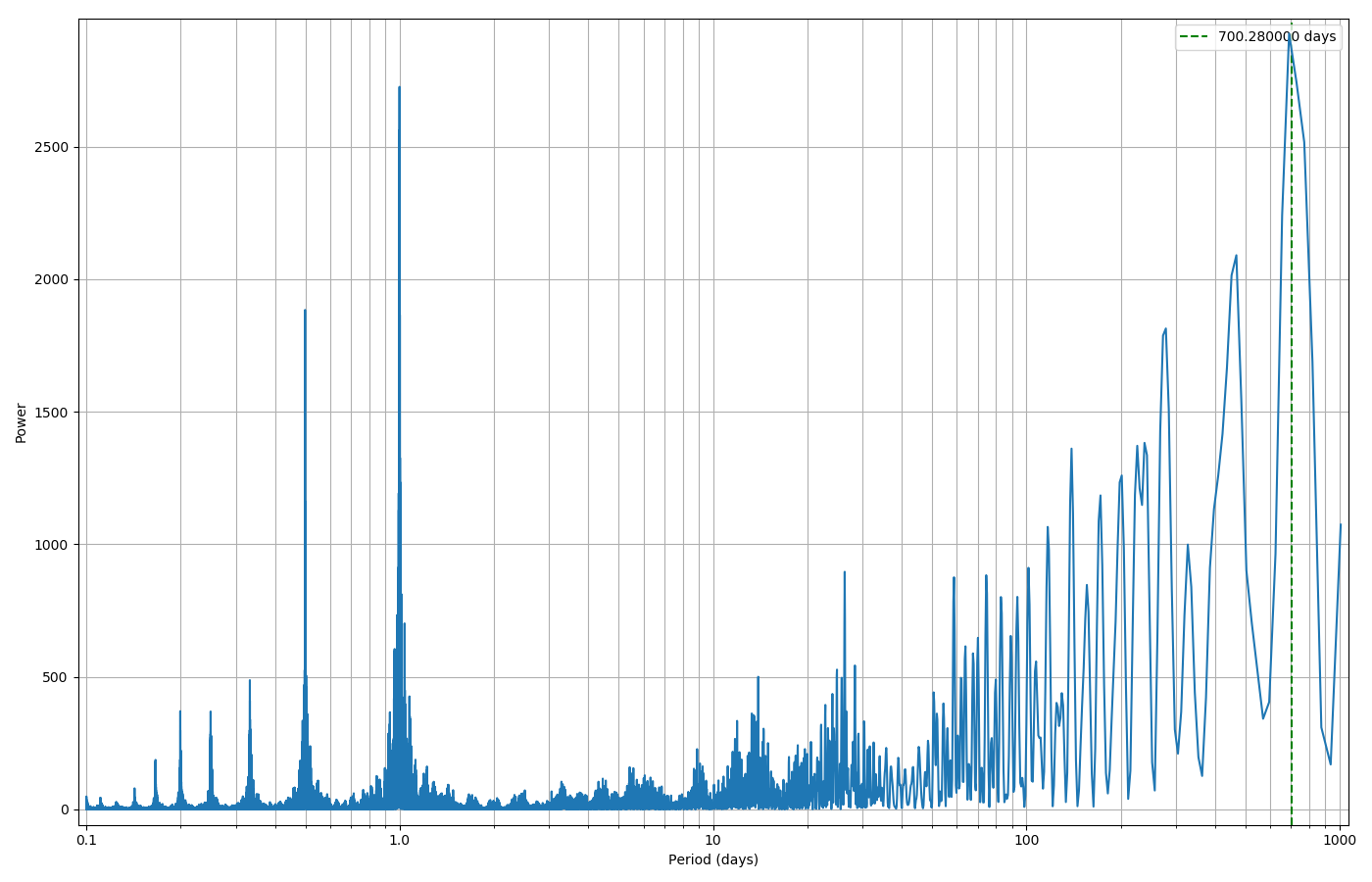}
  \end{center}
  \caption{The Lomb-Scargle periodogram for the saturated pixel fraction in the KELT-North 05 field. The green vertical line marks the period with the greatest power at 700.28 days.}
  \label{fig:periodogram}
\end{figure*}

In order to look for any periodicity in this field, we then run a Lomb-Scargle periodogram on the saturated pixel fraction \citep{Lomb1976, Scargle1982}. As shown in Fig~\ref{fig:periodogram}, we find that the top peak occurs at 700.28 days, with additional smaller peaks corresponding to known astronomical effects, such as the 28-day orbit of the moon. The large peak at 1 day is an aliasing effect caused by the rotation of the Earth (KELT-North only observes at night).

\section{Discussion}
We characterize the observed planet primarily in terms of the orbital characteristics and the resultant incident flux. The relationship between the period and semi-major axis of a planet has been well-characterized, and we utilize that relationship in order to determine the distance of this planet from its host star \citep{Kepler1619}. Using this standard relationship between period and semi-major axis we determine that the semi-major axis of this planet is 1.54 AU around what we infer to be a 1 $M_\odot$ star. We note, however, that due to the brightness of the host star (an apparent magnitude of $\sim$-27) we are unable to use KELT to directly characterize the host star \citep{Torres2010}.

This semi-major axis allows us to roughly characterize the expected incident flux on this planet's surface and the impacts of this flux on its surface and atmosphere. Recent work on solar-mass stars demonstrates that the planet would easily fall within the habitable zone, with estimates of the inner and outer boundaries ranging from 0.99 to 1.68 AU \citep{Kopparapu2013} to 0.95 to 2.4 AU \citep{Ramirez2017}. As the planet is well-suited for follow-up with larger telescopes, and as it is in the habitable zone of its host star, it would be particularly useful to look for signs of water and life on it. For example, it would be conceivable that a higher resolution observation would show tell-tale signs of liquid water on the surface such as seas, or possibly even signs of intelligent life on the planet, such as canals that have been constructed to direct the flow of that water, and would be structures large enough to be observed from great distance. In an interdisciplinary nature, we suggest that it would not be surprising if the structure of these canals is similar to patterns seen in nature, for example, the structure of blood vessels in the human eye, and may appear similar to the figures present in \citet{Retina}.

\section{Summary}

In this paper we have demonstrated how the fraction of saturated pixels can be used to detect the presence of very bright, and very close, planets. In particular, we provide an independent confirmation of a sub-Earth in the habitable zone of a 1 $M_\odot$ star using this technique. The planet we detect is in an orbit with a semi-major axis of 1.54 AU, and the star that it is orbiting is located extremely close to the Earth.

We believe that the close proximity of this planet provides a unique opportunity not just for standard photometric and spectroscopic observations, but also for far more novel approaches at planet characterization. While proposals have been made already for sending probes to stars and planets at greater distances \citep{Hein2017, Heller2017}, these often rely on relatively new and untested techniques. In contrast, the planet we present here can be reached using technology that has been discussed and developed for a substantially longer time \citep{burroughs1917, Dick1966}. Consequentially, it is entirely possible that current technology would be capable of successfully sending probes to provide closer study of this planet.


\bibliographystyle{apalike}
\bibliography{main}

\begin{thebibliography}{}

\bibitem[{Anglada-Escud{\'e}} et~al., 2016]{2016Natur.536..437A}
{Anglada-Escud{\'e}}, G., {Amado}, P.~J., {Barnes}, J., {Berdi{\~n}as}, Z.~M.,
  {Butler}, R.~P., {Coleman}, G.~A.~L., {de La Cueva}, I., {Dreizler}, S.,
  {Endl}, M., {Giesers}, B., {Jeffers}, S.~V., {Jenkins}, J.~S., {Jones},
  H.~R.~A., {Kiraga}, M., {K{\"u}rster}, M., {L{\'o}pez-Gonz{\'a}lez}, M.~J.,
  {Marvin}, C.~J., {Morales}, N., {Morin}, J., {Nelson}, R.~P., {Ortiz}, J.~L.,
  {Ofir}, A., {Paardekooper}, S.-J., {Reiners}, A., {Rodr{\'{\i}}guez}, E.,
  {Rodr{\'{\i}}guez-L{\'o}pez}, C., {Sarmiento}, L.~F., {Strachan}, J.~P.,
  {Tsapras}, Y., {Tuomi}, M., and {Zechmeister}, M. (2016).
\newblock {A terrestrial planet candidate in a temperate orbit around Proxima
  Centauri}.
\newblock {\em \nat}, 536:437--440.

\bibitem[Burroughs, 1917]{burroughs1917}
Burroughs, E. (1917).
\newblock {\em A Princess of Mars}.
\newblock Martian series. Grosset \& Dunlap.

\bibitem[{Delfosse} et~al., 1998]{Delfosse1998}
{Delfosse}, X., {Forveille}, T., {Mayor}, M., {Perrier}, C., {Naef}, D., and
  {Queloz}, D. (1998).
\newblock {The closest extrasolar planet. A giant planet around the M4 dwarf GL
  876}.
\newblock {\em \aap}, 338:L67--L70.

\bibitem[Dick, 1966]{Dick1966}
Dick, P.~K. (1966).
\newblock We can remember it for you wholesale.
\newblock {\em The Magazine of Fantasy \& Science Fiction}.

\bibitem[{Dressing} and {Charbonneau}, 2015]{Dressing2015}
{Dressing}, C.~D. and {Charbonneau}, D. (2015).
\newblock {The Occurrence of Potentially Habitable Planets Orbiting M Dwarfs
  Estimated from the Full Kepler Dataset and an Empirical Measurement of the
  Detection Sensitivity}.
\newblock {\em \apj}, 807:45.

\bibitem[{Dunsby}, 2018a]{2018ATel11449....1D}
{Dunsby}, P. (2018a).
\newblock {Erratum to ATel \#11448: Very bright optical transient near the
  Trifid and Lagoon Nebulae}.
\newblock {\em The Astronomer's Telegram}, 11449.

\bibitem[{Dunsby}, 2018b]{2018ATel11448....1D}
{Dunsby}, P. (2018b).
\newblock {Very bright optical transient near the Trifid and Lagoon Nebulae}.
\newblock {\em The Astronomer's Telegram}, 11448.

\bibitem[{Hein} et~al., 2017]{Hein2017}
{Hein}, A.~M., {Long}, K.~F., {Fries}, D., {Perakis}, N., {Genovese}, A.,
  {Zeidler}, S., {Langer}, M., {Osborne}, R., {Swinney}, R., {Davies}, J.,
  {Cress}, B., {Casson}, M., {Mann}, A., and {Armstrong}, R. (2017).
\newblock {The Andromeda Study: A Femto-Spacecraft Mission to Alpha Centauri}.
\newblock {\em ArXiv e-prints}.

\bibitem[{Heller} et~al., 2017]{Heller2017}
{Heller}, R., {Hippke}, M., and {Kervella}, P. (2017).
\newblock {Optimized Trajectories to the Nearest Stars Using Lightweight
  High-velocity Photon Sails}.
\newblock {\em \aj}, 154:115.

\bibitem[{Kaba} et~al., 2014]{Retina}
{Kaba}, D., {Wang}, C., {Li}, Y., {Salazar-Gonzalez}, A., {Liu}, X., and
  {Serag}, A. (2014).
\newblock {Retinal blood vessels extraction using probabilistic modelling}.
\newblock {\em Health Information Science and Systems}, 2.

\bibitem[{Kalas} et~al., 2008]{Kalas2008}
{Kalas}, P., {Graham}, J.~R., {Chiang}, E., {Fitzgerald}, M.~P., {Clampin}, M.,
  {Kite}, E.~S., {Stapelfeldt}, K., {Marois}, C., and {Krist}, J. (2008).
\newblock {Optical Images of an Exosolar Planet 25 Light-Years from Earth}.
\newblock {\em Science}, 322:1345.

\bibitem[{Kepler}, 1619]{Kepler1619}
{Kepler}, J. (1619).
\newblock {\em {Ioannis Keppleri harmonices mundi libri V : quorum primus
  harmonicus ... quartus metaphysicus, psychologicus et astrologicus
  geometricus ... secundus architectonicus ... tertius proprie ... quintus
  astronomicus {\&} metaphysicus ... : appendix habet comparationem huius
  operis cum harmonices Cl. Ptolemaei libro III cumque Roberti de Fluctibus ...
  speculationibus harmonicis, operi de macrocosmo {\&} microcosmo insertis}}.

\bibitem[{Kopparapu} et~al., 2013]{Kopparapu2013}
{Kopparapu}, R.~K., {Ramirez}, R., {Kasting}, J.~F., {Eymet}, V., {Robinson},
  T.~D., {Mahadevan}, S., {Terrien}, R.~C., {Domagal-Goldman}, S., {Meadows},
  V., and {Deshpande}, R. (2013).
\newblock {Habitable Zones around Main-sequence Stars: New Estimates}.
\newblock {\em \apj}, 765:131.

\bibitem[{Lomb}, 1976]{Lomb1976}
{Lomb}, N.~R. (1976).
\newblock {Least-squares frequency analysis of unequally spaced data}.
\newblock {\em \apss}, 39:447--462.

\bibitem[{Marcy} et~al., 1998]{Marcy1998}
{Marcy}, G.~W., {Butler}, R.~P., {Vogt}, S.~S., {Fischer}, D., and {Lissauer},
  J.~J. (1998).
\newblock {A Planetary Companion to a Nearby M4 Dwarf, Gliese 876}.
\newblock {\em \apjl}, 505:L147--L149.

\bibitem[{Pepper} et~al., 2012]{Pepper2012}
{Pepper}, J., {Kuhn}, R.~B., {Siverd}, R., {James}, D., and {Stassun}, K.
  (2012).
\newblock {The KELT-South Telescope}.
\newblock {\em \pasp}, 124:230.

\bibitem[{Pepper} et~al., 2007]{Pepper2007}
{Pepper}, J., {Pogge}, R.~W., {DePoy}, D.~L., {Marshall}, J.~L., {Stanek},
  K.~Z., {Stutz}, A.~M., {Poindexter}, S., {Siverd}, R., {O'Brien}, T.~P.,
  {Trueblood}, M., and {Trueblood}, P. (2007).
\newblock {The Kilodegree Extremely Little Telescope (KELT): A Small Robotic
  Telescope for Large-Area Synoptic Surveys}.
\newblock {\em \pasp}, 119:923--935.

\bibitem[{Ramirez} and {Kaltenegger}, 2017]{Ramirez2017}
{Ramirez}, R.~M. and {Kaltenegger}, L. (2017).
\newblock {A Volcanic Hydrogen Habitable Zone}.
\newblock {\em \apjl}, 837:L4.

\bibitem[{Scargle}, 1982]{Scargle1982}
{Scargle}, J.~D. (1982).
\newblock {Studies in astronomical time series analysis. II - Statistical
  aspects of spectral analysis of unevenly spaced data}.
\newblock {\em \apj}, 263:835--853.

\bibitem[{Siverd} et~al., 2012]{Siverd2012}
{Siverd}, R.~J., {Beatty}, T.~G., {Pepper}, J., {Eastman}, J.~D., {Collins},
  K., {Bieryla}, A., {Latham}, D.~W., {Buchhave}, L.~A., {Jensen}, E.~L.~N.,
  {Crepp}, J.~R., {Street}, R., {Stassun}, K.~G., {Gaudi}, B.~S., {Berlind},
  P., {Calkins}, M.~L., {DePoy}, D.~L., {Esquerdo}, G.~A., {Fulton}, B.~J.,
  {F{\H u}r{\'e}sz}, G., {Geary}, J.~C., {Gould}, A., {Hebb}, L., {Kielkopf},
  J.~F., {Marshall}, J.~L., {Pogge}, R., {Stanek}, K.~Z., {Stefanik}, R.~P.,
  {Szentgyorgyi}, A.~H., {Trueblood}, M., {Trueblood}, P., {Stutz}, A.~M., and
  {van Saders}, J.~L. (2012).
\newblock {KELT-1b: A Strongly Irradiated, Highly Inflated, Short Period, 27
  Jupiter-mass Companion Transiting a Mid-F Star}.
\newblock {\em \apj}, 761:123.

\bibitem[{Torres}, 2010]{Torres2010}
{Torres}, G. (2010).
\newblock {On the Use of Empirical Bolometric Corrections for Stars}.
\newblock {\em \aj}, 140:1158--1162.

\end{thebibliography}


\end{multicols}

\end{document}